\documentclass[10pt,twocolumn,aps,pra,longbibliography,floatfix,%
reprint,superscriptaddress]{revtex4-2}

\def\nz{{\sf NZ}}
\usepackage[T2A]{fontenc}
\usepackage[russian,english]{babel}
\usepackage[utf8]{inputenc}
\def\nz{{\sf ИZ}}

\usepackage{amsmath,amssymb,amsfonts}
\usepackage{graphicx}
\graphicspath{{./}{./pdf/}}
\usepackage{xcolor}

\usepackage{braket}

\usepackage{hyperref}
\hypersetup{
    colorlinks=true,
    linkcolor=blue,
    citecolor=blue,
    filecolor=magenta,
    urlcolor=cyan
}

\usepackage{qcircuit}
\let\qqw=\qw
\renewcommand{\qw}{}
\let\oldgate=\gate

\renewcommand{\gate}[1]{\oldgate{#1}\qqw}
\let\oldctrl=\ctrl
\renewcommand{\ctrl}[1]{\oldctrl{#1}\qqw}
\let\oldtarg=\targ
\newcommand{\ctarg}{\oldtarg\cw}
\renewcommand{\targ}{\oldtarg\qqw}

\begin{document}

\title{Optimized noise-resilient surface code teleportation interfaces}

\author{Mohamed A. Shalby}
\affiliation{Department of Physics and Astronomy, University of California, Riverside 92521 USA}

\author{Renyu Wang}
\affiliation{Department of Physics and Astronomy, University of California, Riverside 92521 USA}

\author{Denis Sedov}
\affiliation{Institute for Theoretical Physics III, University of Stuttgart, 70550 Stuttgart Germany}

\author{Leonid P. Pryadko}
\affiliation{Department of Physics and Astronomy, University of California, Riverside 92521 USA}

\date{\today}

\begin{abstract}
  Connecting two surface‐code patches may require significantly higher
  noise at the interface. We show, via circuit‐level simulations under
  a depolarizing noise model with idle errors, that surface codes
  remain fault tolerant despite substantially elevated interface error
  rates. Specifically, we compare three strategies---direct noisy links,
  gate teleportation, and a CAT‐state gadget---for both rotated and
  unrotated surface codes, and demonstrate that careful design can
  mitigate hook errors in each case so that the full code distance is
  preserved for both $X$ and $Z$. Although these methods differ in
  space and time overhead and performance, each offers a viable route
  to modular surface‐code architectures. Our results, obtained with
  {\tt Stim} and {\tt PyMatching}, confirm that high‐noise
  interfaces can be integrated fault‐tolerantly without compromising
  the code's essential properties, indicating that fault-tolerant
  scaling of error-corrected modular devices is within reach with
  current technology.
\end{abstract}

\maketitle

\noindent{\bf Introduction.}
Quantum error correction (QEC) is crucial for achieving large-scale
fault-tolerant quantum computation~\cite{Shor-FT-1996}. Surface
codes, in particular, offer a compelling balance of high thresholds,
planar connectivity, and relative architectural
simplicity~\cite{kitaev-anyons,Dennis-Kitaev-Landahl-Preskill-2002,%
  Fowler-Mariantoni-Martinis-Cleland-2012,Bravyi-Kitaev-1998}.

However, scaling up to large devices poses fabrication, control, and
thermal challenges. As a result, there is a growing interest in
\emph{modular} quantum processors, where smaller processing modules
are interconnected to form a larger, fault-tolerant computing
network~\cite{Monroe-etal-modular-2014,Nickerson2014}. A similar
approach is also needed in quantum communication to transfer quantum
data between different stations at a distance, or in quantum memory to
have higher storage capacity~\cite{Muralidharan2016}.  In such 
systems, the \emph{interfaces} between two modules often exhibit
elevated noise compared to the bulk, necessitating special boundary
gadgets or coupling strategies to maintain global fault
tolerance~\cite{Ramette-Sinclair-Breuckmann-Vuletic-2023}.

Within phenomenological error model, a map to a spin model indicates
that a modular quantum computer with modules in the form of
sufficiently large square patches of surface code can be below the
threshold whenever bulk qubits are, and with any boundary error
probability $q<1/2$ \cite{Sedov-Wang-Pryadko-APS-2023}.  More accurate
mean-field analysis and circuit-level simulations suggest that error
rates for gates across the boundary up to roughly an order of
magnitude higher than the threshold required for the bulk operations
within each module can be
tolerated\cite{Ramette-Sinclair-Breuckmann-Vuletic-2023,Sedov-Wang-Pryadko-APS-2023}.
This indicates that scalable quantum computing based on modules
connected with noisy inter-modular links (whether optical,
superconducting, or ion‐based) may be feasible with current or near-term
technology.

On this front, various technologies are being explored to connect
separated quantum modules. For superconducting qubits, waveguide-based
couplers~\cite{Magnard2020,CampagneIbarcq2020} or 3D-coaxial
cables~\cite{Rosenberg2020} have shown promise as inter-chip links but
can introduce additional loss and decoherence. In photonic
architectures, fiber or integrated waveguide channels can mediate
entangled links~\cite{Main2025,Pelc2011,Li2022Photonics}. Similarly,
trapped-ion platforms use photonic interconnects or phonon-bus
couplings~\cite{Monroe-etal-ion-2021,Stephenson2020} to bridge
multiple ion traps.  Such connectors, whether based on stationary or
flying qubits, may require additional boundary gadgets, e.g.,
teleportation circuits.  An important question is how can such
interface gadgets be integrated with the surface code measurement
cycle for qubits in the bulk.

\begin{figure}[htbp]
  \centering
  \begin{tabular}{lll}
  (a)\quad\strut & 
  \begin{minipage}{0.4\linewidth}
    \centering
    \Qcircuit @C=0.5em @R=0.5em {
      &&& \qqw                 & \qqw    &\ctrl{2}    & \qqw&\qqw\\
      &&&&&\textcolor{red}{--------}&\\
      &&& \lstick{\ket{0}}     &\targ  & \targ  & \gate{M_Z} & \cw \\
      &&&\qqw                   & \ctrl{-1}    & \qqw & \qqw &\qqw\\
    }
  \end{minipage}
  & 
  \begin{minipage}{0.4\linewidth}
    \centering
    \Qcircuit @C=0.5em @R=0.5em {
      &&& \qqw                 & \qqw    &\targ    & \qqw&\qqw\\
      &&&&&\textcolor{red}{--------}&\\
      &&& \lstick{\ket{+}}     &\ctrl{1}  & \ctrl{-2}  & \gate{M_X} & \cw \\
      &&&\qqw                   & \targ    & \qqw & \qqw &\qqw\\
    }
  \end{minipage}\hfill\\[3em]
  (b) &
  \begin{minipage}{0.4\linewidth}
    \centering
    \Qcircuit @C=0.5em @R=0.5em {
      &&& \qqw                 & \qqw    &\qqw   & \ctrl{1}  & \qqw & \qqw&\qqw\\
      &&& \lstick{\ket{+}}     &\ctrl{2}  &\qqw       &\targ&\qqw& \gate{M_Z} &\cctrl{2} \\
      &&&&&\textcolor{red}{--------}&\\
      &&& \lstick{\ket{0}}     &\targ    & \targ  &\qqw&\qqw& \gate{M_Z} & \ctarg&\cw \\
      &&&\qqw                   & \qqw    &\ctrl{-1}& \qqw & \qqw & \qqw &\qqw\\
    }
  \end{minipage}
   &
  \begin{minipage}{0.4\linewidth}
    \centering
    \Qcircuit @C=0.5em @R=0.5em {
      & \qqw                 & \qqw    &\qqw   & \targ  & \qqw & \qqw&\qqw\\
      & \lstick{\ket{+}}     &\ctrl{2}  &\qqw       &\ctrl{-1}&\qqw& \gate{M_X} & \cctrl{2} \\
      &&&&\textcolor{red}{--------}&\\
      & \lstick{\ket{0}}     &\targ    & \ctrl{1}  &\qqw&\qqw& \gate{M_X} & \ctarg&\cw \\
      &\qqw                   & \qqw    &\targ & \qqw & \qqw & \qqw &\qqw\\
    }
  \end{minipage}
  \\[3em]
  {(c)} &
  \begin{minipage}{0.4\linewidth}
    \centering
    \Qcircuit @C=0.5em @R=0.5em {
      && \qqw                 & \qqw       &\qqw        & \ctrl{0}  & \qqw & \qqw&\qqw&\qqw\\
      && \lstick{\ket{+}}     &\ctrl{1}  &\qqw            &\ctrl{-1}&\qqw& \gate{M_X} & \cctrl{3} \\
      && \lstick{\ket{0}}     &\targ     & \ctrl{2}  &\qqw&\qqw& \gate{M_X} & \cw&\cctrl{2} \\
      &&&&&\textcolor{red}{--------}&\\
      && \lstick{\ket{0}}     &\qqw     &\targ& \ctrl{1}  &\qqw& \gate{M_X} & \ctarg & \ctarg&\cw \\
      &&\qqw                   & \qqw      &\qqw&\ctrl{0} & \qqw & \qqw & \qqw&\qqw \\
    }
  \end{minipage}\qquad\strut%
  &
  \begin{minipage}{0.4\linewidth}
    \centering
    \Qcircuit @C=0.5em @R=0.5em {
      & \qqw                 & \qqw       &\qqw        & \targ  & \qqw & \qqw&\qqw&\qqw\\
      & \lstick{\ket{+}}     &\ctrl{1}  &\qqw            &\ctrl{-1}&\qqw& \gate{M_x} & \cctrl{3} \\
      & \lstick{\ket{0}}     &\targ     & \ctrl{2}  &\qqw&\qqw& \gate{M_X} &\cw& \cctrl{2} \\
      &&&&\textcolor{red}{--------}&\\
      & \lstick{\ket{0}}     &\qqw     &\targ& \ctrl{1}  &\qqw& \gate{M_X} & \ctarg& \ctarg&\cw\\
      &\qqw                   & \qqw      &\qqw&\targ & \qqw & \qqw & \qqw&\qqw \\
    }
  \end{minipage}%
  \end{tabular}
  \caption{(Color online) (a) DL measurement for two data qubits across
    the boundary.  Left: $ZZ$ product; Right: $XX$ product. (b) same for CAT
    gadget. (c) same for GT gadget.  Red dashed lines indicate the
    interface location.}
  \label{fig:gadgets}
\end{figure}
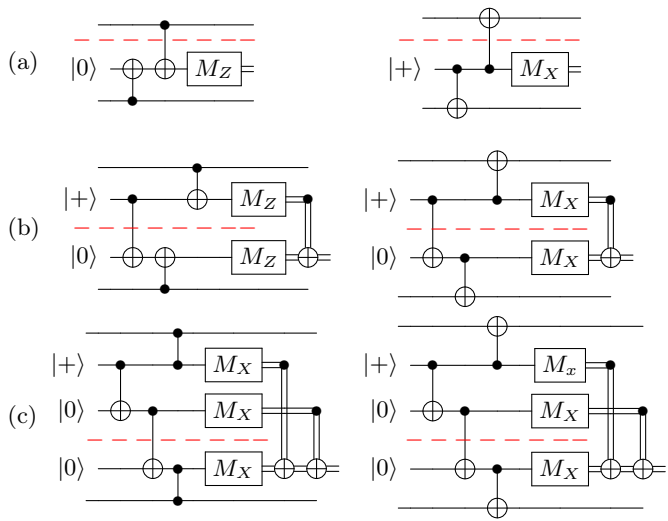

The goal of this work is to compare the performance of several
interface gadgets under generic circuit noise, with fixed
boundary-to-bulk error rate ratios $\gamma\equiv q/p$.  Specifically,
we construct heavily optimized circuits for two such methods,
\emph{cat state} (CAT) gadget~\cite{Huang2021} and \emph{gate
  teleportation} (GT) gadget~\cite{Gottesman1999}, as well as a
simpler noisy direct-link (DL) interface [Fig.~\ref{fig:gadgets}],
with both \emph{rotated} and \emph{unrotated} surface codes.  For all
six boundary configurations, using error rate ratios
$\gamma\in \{1,10\}$, we calculate (pseudo)thresholds and values of
$\Lambda$ at fixed $p=0.345\%$.  We conclude that all these methods
can be used to connect patches of surface codes, with reasonably close
(pseudo)thresholds and below-threshold behavior.  In particular, our
results validate approximations made in
Ref.~\cite{Ramette-Sinclair-Breuckmann-Vuletic-2023} for the case of
gate teleportation gadgets connecting two unrotated surface code
patches.

\noindent{\bf Circuits:} In the bulk, we adopt the standard \nz\
surface code measurement schedule\cite{Tomita-Svore-2014}.  This
ensures that weight-two hook errors are aligned across logical
operators, thus ensuring that any fault causes no more damage than a
single-qubit error should.

For DL gadget, the same schedule is also adopted at the boundary, with
the error rate $q=\gamma p$ for the two-qubit gates crossing the
boundary indicated with a red dashed line.  Here $p$ is the error rate
in the bulk of both patches, and $\gamma$ is the boundary-to-bulk
error rate ratio.  Two-qubit versions of such a single-ancilla
measurement circuit, for measuring $ZZ$ and $XX$, are shown in
Fig.~\ref{fig:gadgets}(a), for easier comparison with the other two
gadgets.

Circuit diagrams for the CAT and GT gadgets, respectively, are shown
in Fig.~\ref{fig:gadgets}(b) and (c).  Again, each gadget is shown in
two variants to accommodate either a $Z$-type or an $X$-type
stabilizer generator.  For clarity, we focus on coupling only two
qubits residing on separate patches.  Any CNOT gate crossing this line
is treated as a high-noise, long-range operation, thereby introducing
elevated error rates at the interface.

The same gadgets can couple multiple qubits on both sides as required,
depending on the code type, stabilizer weight, measurement schedule,
and interface geometry. Also, each of these syndrome-extraction
circuits can be remapped into alternate but logically equivalent
forms~\cite{Barenco-1995}. Such transformations can be particularly
advantageous for specific hardware platforms, where gate availability,
operation times, or parallelization opportunities may differ
significantly.

While DL measurement does not affect the depth of the measurement
circuit, CAT and GT gadgets require circuits with 5 CNOT gates applied
sequentially (circuit depth 5).  To keep these circuits in sync with
measurements in the bulk, we insert an additional idle step, thus also
increasing the depth of the bulk measurement circuit to 5.

Second challenge in designing circuits for CAT and GT boundary
gadgets is to suppress the hook errors.  Although partial remedies
(e.g., reversal of measurement schedules between
rounds~\cite{Debroy2024}) can limit the damage, our focus here is on
configurations that maintain the full code distance for both logical
observables.

Damaging hook errors appear when a stabilizer generator of weight four
is split evenly between patches, effectively creating two pairs of
data qubits.  In this case, hook errors are oriented along the
boundary, necessarily reducing the effective distance with respect to
either logical $X$ or logical $Z$ observable.

\begin{figure}[htbp]
	\centering
	\begin{minipage}{0.3\columnwidth}
		\centering
		\includegraphics[width=\linewidth]{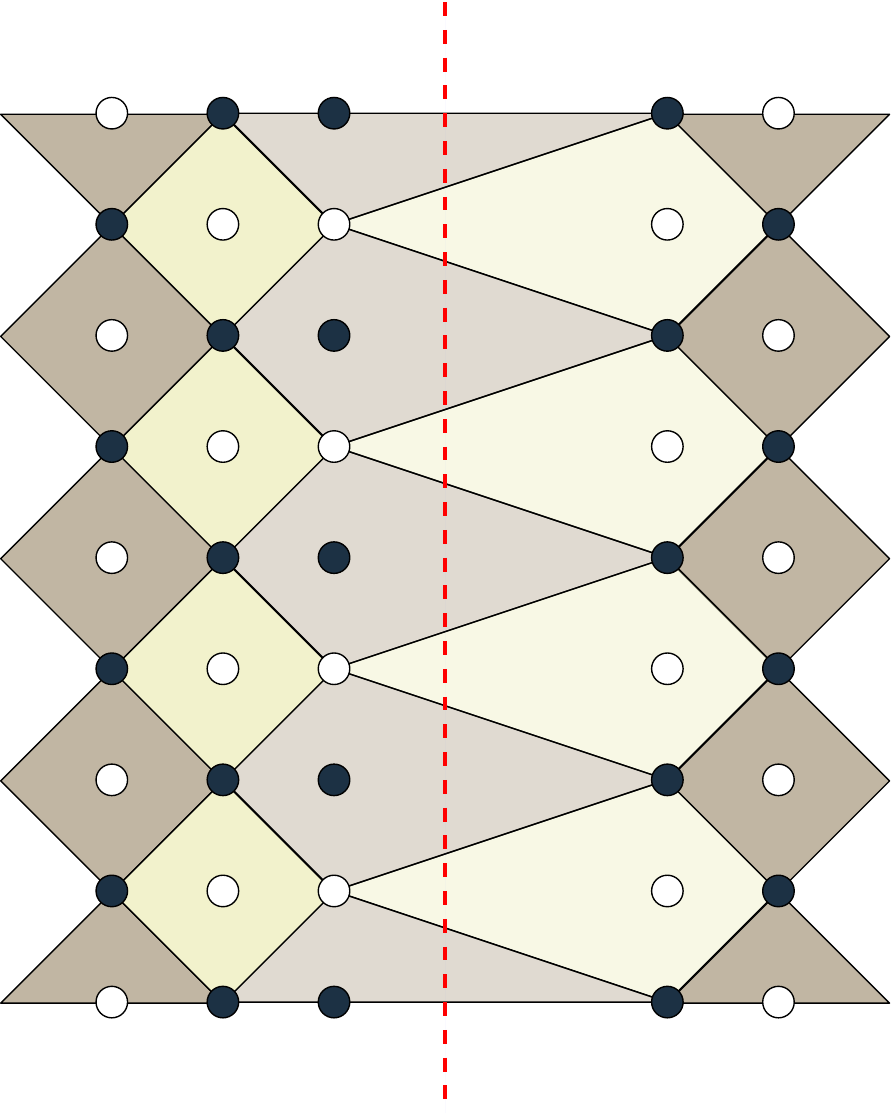}
		{\footnotesize (a)}
	\end{minipage}
	\hfill
	\begin{minipage}{0.3\columnwidth}
		\centering
		\includegraphics[width=\linewidth]{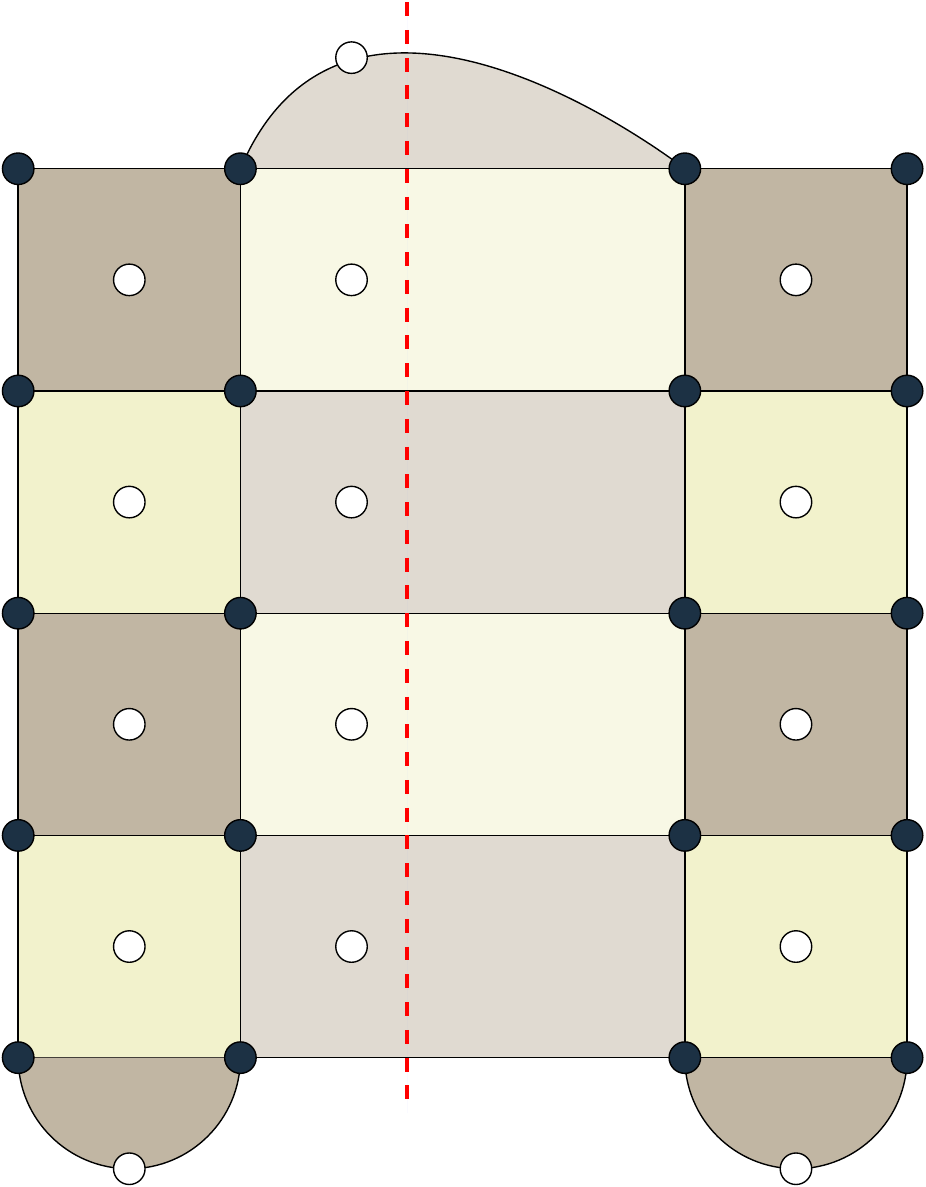}
		{\footnotesize (b)}
	\end{minipage}
	\hfill
	\begin{minipage}{0.3\columnwidth}
		\centering
		\includegraphics[width=\linewidth]{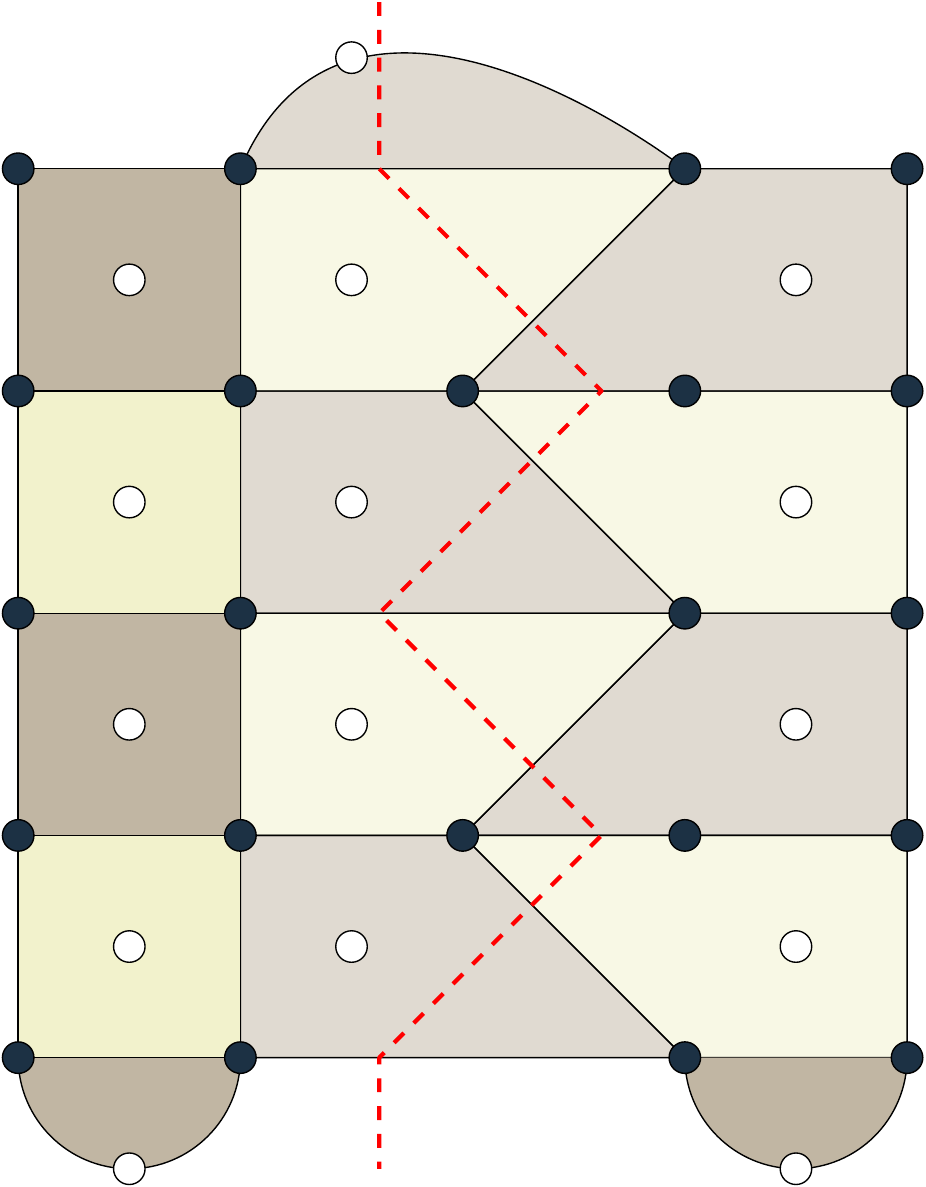}
		{\footnotesize (c)}
	\end{minipage}
	
	\caption{(Color online)  Different surface-code
          interface configurations. Panel (a) shows two patches of
          unrotated surface code with a straight-line interface, panel
          (b) depicts a rotated surface code with a straight-line
          interface, and panel (c) shows a rotated surface code with a
          zig-zag interface.}
	\label{fig:surface_code_layouts}
\end{figure}

In unrotated surface codes, boundary stabilizer generators typically
encompass three data qubits on one side and only one on the other
[Fig.\ref{fig:surface_code_layouts}(a)], and hook errors can be
controlled just by choosing the addressing schedules.  However, in
rotated surface codes, a linear boundary yields a 2–2 split
[Fig.\ref{fig:surface_code_layouts}(b)], prompting us to adopt a
zigzag boundary geometry
[Fig.~\ref{fig:surface_code_layouts}(c)]. This design ensures no
stabilizer generator is split evenly between the two patches,
eliminating critical hook errors at the interface.

\begin{figure*}[htbp]
  \centering
  \includegraphics[width=0.4\textwidth]{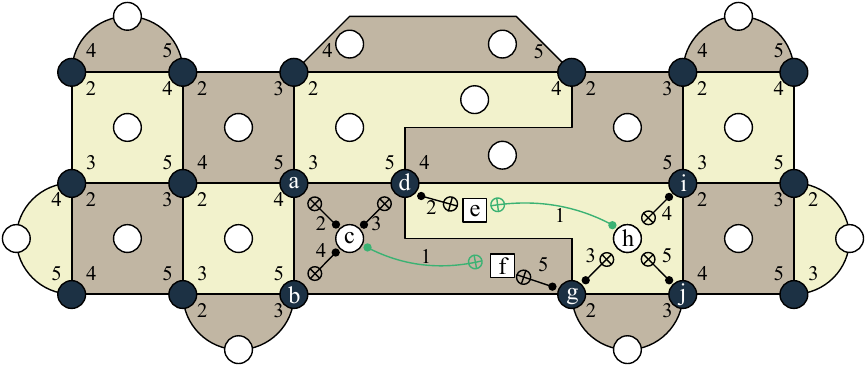} \qquad 
  \includegraphics[width=0.4\textwidth]{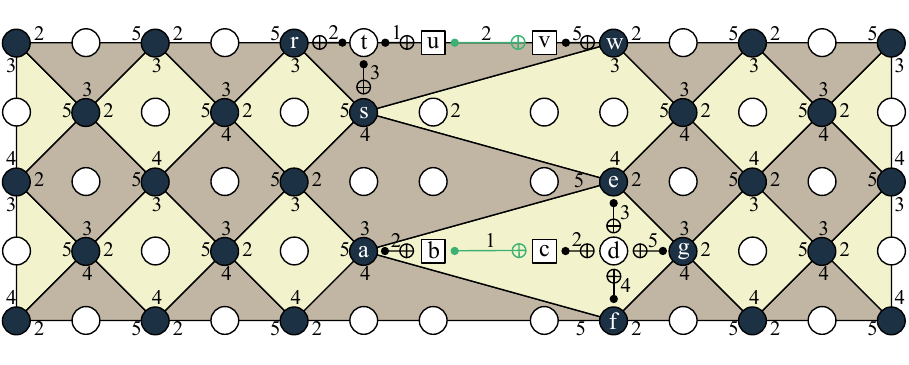} 
  \caption{(Color online) Detailed layouts and addressing schedules
    (numbers) for rotated surface code with CAT gadget at the
    interface (Left) and unrotated surface code with gate
    teleportation (Right). Data qubits are represented with dark
    circles, ancillary qubits with white circles, while extra qubits
    are represented using white squares. CNOT gates with elevated
    noise are colored green. More details in the Appendix}
  \label{fig:two-layouts}
\end{figure*}

In Fig.~\ref{fig:two-layouts}, as a representative sample, we show
detailed layouts and addressing schedules for two distance-3
interfaces, rotated surface code with CAT gadget at the interface
(Left) and unrotated surface code with gate teleportation (Right).
The modified boundary design for rotated surface codes effectively
avoids critical hook errors.  These, as well as all other circuits we
designed, can be implemented in a single-layer planar qubit layout.

The addressing schedule is shown in Fig.~\ref{fig:two-layouts} by the
numbers which indicate the time step for addressing a particular data
qubit or executing a particular CNOT gate.  In the bulk, this is the
familiar \nz\ pattern, augmented by an additional initial step that
accommodates gadget-specific requirements.  During this added step,
all other ancillas and data qubits remain idle and thus experience
idling noise. This design ensures that circuit modifications at the
interface preserve the overall fault tolerance of the system.

{\bf Error model:} We used a standard circuit-level depolarizing noise
model with idle errors, subject to four key assumptions. First,
single-qubit gate errors occur after every single-qubit operation with
probability \(p_1\). Second, two-qubit gate errors arise after each
entangling operation (e.g., CNOT) with probability \(p_2\). Third,
qubits that are not actively involved in an operation are exposed to
idling errors, modeled as single-qubit depolarizing noise with
probability \(p_{\mathrm{idle}}\). Finally, initialization and
measurement errors occur with probability \(p_{\mathrm{flip}}\),
manifesting as an $X$-type error for a $Z$-basis measurement (or a
$Z$-type error for an $X$-basis measurement) both immediately after
initialization and just before measurement.

In our simulations, we set
\(p_1 = p_2 = p_{\mathrm{idle}} = p_{\mathrm{flip}}\). Within the
\emph{bulk} of each surface-code patch, these error probabilities are
collectively denoted by \(p\). By contrast, at the \emph{interface},
the two-qubit gate error rate, \(p_2\), is elevated by a factor
\(\gamma\), so \(q = \gamma \, p\). We primarily focus on the case of
\(\gamma = 10\), which highlights scenarios with significantly higher
boundary noise, while \(\gamma = 1\) serves as a baseline to isolate
the effects of interface gadgets.

{\bf Simulations:} We simulated the described circuits with {\tt Stim}\cite{Gidney-2021-stim}.
Specifically, we implemented interfaces between a pair of square
distance-$d$ patches of surface code, for odd distances in the range
$3\le d\le 11$. Two variants of each simulation were constructed, with
data qubits initially prepared in $\ket0$ or $\ket+$ states, and the
final data-qubit measurement done in the $Z$ or $X$ basis,
respectively.  Each circuit included initial data qubit preparation,
$N=d$ identical measurement rounds, followed by a final round of data
qubit measurements.  In addition, one- and two-qubit Pauli errors were
inserted at the end of every time step as required by the error model.

The complete {\tt Stim} simulation also includes the annotations for
\emph{detector events} representing known relations between
measurements in the absence of errors, and the \emph{observables}
which are used to check whether the decoding was correct.  For
example, the detector events could be the differences between
measurement results in the subsequent rounds of stabilizer
measurements, or similar relations between the $X$- or $Z$-stabilizer
measurements in the last round and the final data-qubit measurements.
Similarly, the observables correspond to the $X$- or $Z$-logical
operators whose expected values in the absence of errors can be
deduced from the data qubit initialization and the final measurement
results.

The constructed circuits were simulated on a laptop computer with {\tt
  Stim}~\cite{Gidney-2021-stim}, using {\tt PyMatching}
\cite{Higgott-2021} decoder.  For each point, the maximum number of
samples was set to $3\times 10^6$, with termination after
$3\times 10^3$ logical errors.

\begin{figure}[htbp]
	\centering
	\includegraphics[width=\columnwidth]{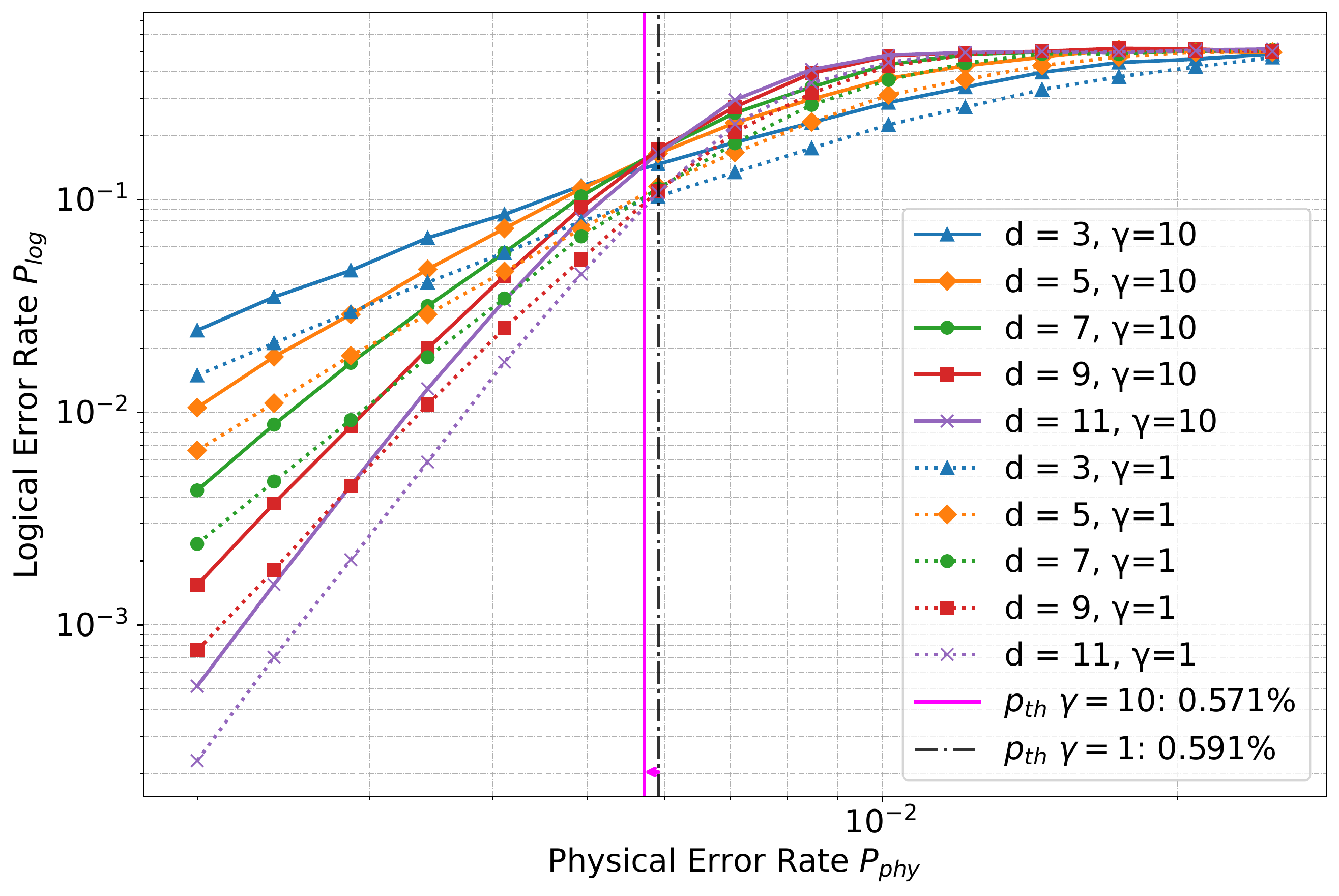}
	\caption{(Color online) Logical error vs.\ physical error
          rates for rotated surface code patches connected with CAT
          gadgets. Solid lines are used for $\gamma=10$, dashed lines
          for $\gamma=1$. The logical observable being measured is
          \emph{across} the interface.  Vertical lines indicate the
          positions of the crossing points.}
	\label{fig:Rotated_CAT_Across_x10}
\end{figure}

{\bf Simulation results} for each family of circuits (varying by
interface gadget, code orientation, and the observable measured) and
for each value of the interface noise factor $\gamma$, are presented
in threshold-style plots, where the logical error rate $p_L$ is
plotted as a function of the error parameter $p$, for the odd patch
distances $3\le d\le 11$.  Such plots for the two boundary
configurations in Fig.~\ref{fig:two-layouts} are shown in
Figs.~\ref{fig:Rotated_CAT_Across_x10} and
\ref{fig:unrotated_Direct_Across_x10}.  (The complete set of plots is
given in the Appendix.)

\begin{figure}[htbp]
	\centering
	\includegraphics[width=\columnwidth]{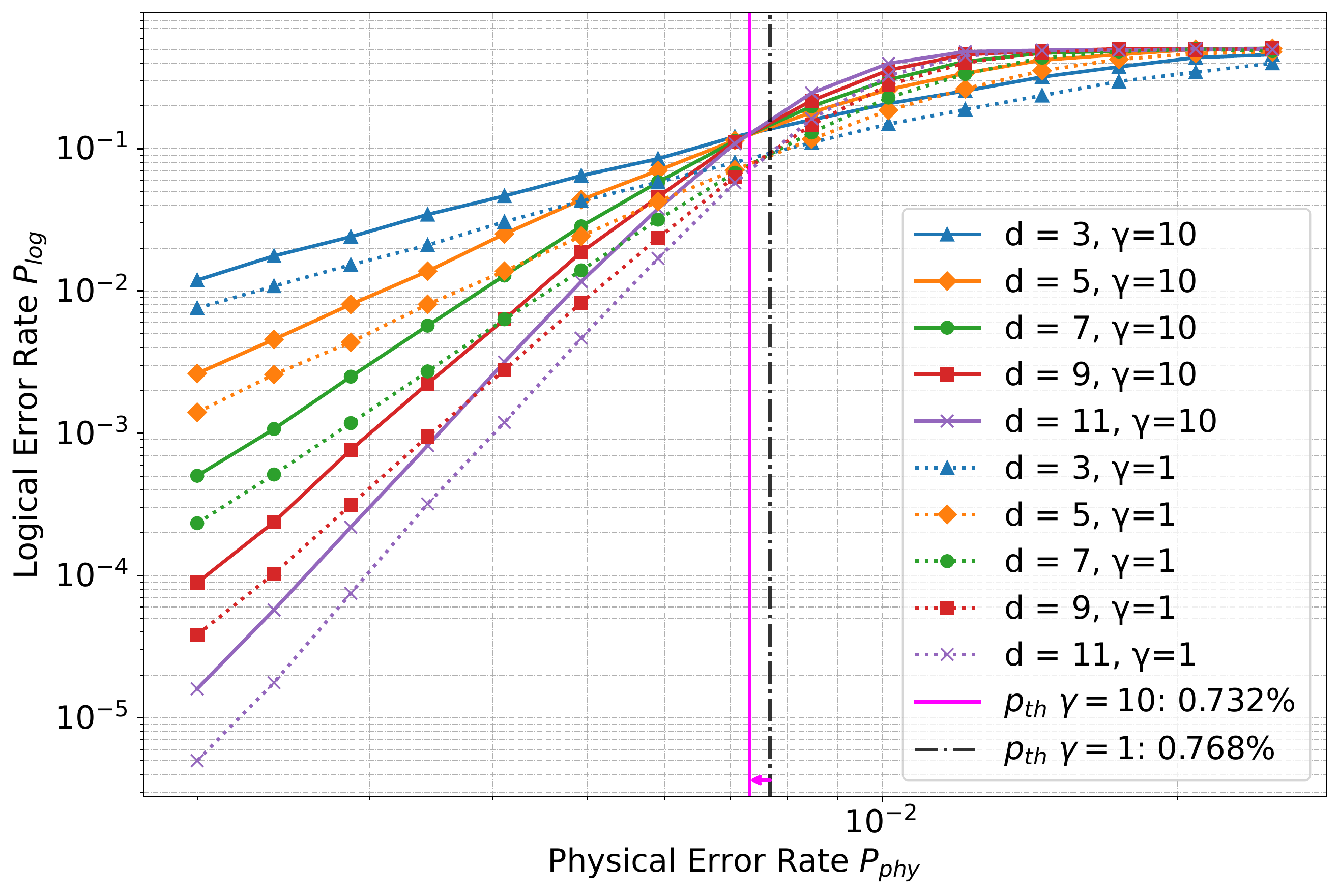}
	\caption{(Color online) Same as in
          Fig.~\ref{fig:Rotated_CAT_Across_x10} but for the DL gadget.}
	\label{fig:unrotated_Direct_Across_x10}
\end{figure}

At smaller values of $p$, all curves for codes with distance $d=2t+1$
have the logarithmic slope approximately equal $t+1$ (asymptotic form
$\propto p^{t+1}$), in agreement with the expected circuit distances
(we only were able to explicitly verify the circuit distances for
$d\le 7$).

The curves for different distances show reasonably good crossing
points, with logical error rates decreasing (increasing) with the
distance, respectively, to the left (to the right) of the crossing
point.  Generally, a threshold is defined as the value of the physical
error rate \(p\) at which the logical error rate \(p_L\) stops
improving with increasing code distance \(d\), in the limit of large
distances.  In a finite-size simulation, an abscissa of a crossing
point gives a pseudothreshold, a finite-size approximation to the
threshold.  The pseudothreshold values $p_{\rm th}(\gamma)$ computed by 
fitting the data close to the crossing points are collected in
Tab.~\ref{tab:combined_results}.

In addition, in Tab.~\ref{tab:combined_results} we also give the
lambda-factors $ \Lambda_{5/7}(\gamma)$, computed at the fixed
physical error rate \(p = 3.45 \times 10^{-3}\) as a ratio of logical
error rates for $d=5$ and $d=7$.  This parameter quantifies how
strongly the logical error rate is suppressed with increasing code
distance.  In general,
\[
\Lambda_{d/(d+2)} \;=\; {p_L^{(d)}}/{p_L^{(d+2)}},
\]
so larger \(\Lambda\) values correspond to more effective
distance-based error suppression.

\renewcommand{\arraystretch}{1.1}
\begin{table*}
  \centering
  \begin{tabular}{c c c | l l l l | l l l l }\hline
    \textbf{Code Type} & ~\textbf{Gadget} & ~\textbf{Depth}
    & \multicolumn{4}{c|}{\textbf{Parallel}} 
    & \multicolumn{4}{c}{\textbf{Across}}\\    \hline
                       & && \multicolumn{1}{c}{\(p_{\rm th}(1)\)}
    & \multicolumn{1}{c}{\(p_{\rm th}(10)\)} &\multicolumn{1}{c}{\(\Lambda_{5/7}(1)\)}
                       & \multicolumn{1}{c|}{\(\Lambda_{5/7}(10)\)}
      &\multicolumn{1}{c}{\(p_{\rm th}(1)\)}
    & \multicolumn{1}{c}{\(p_{\rm th}(10)\)} &\multicolumn{1}{c}{\(\Lambda_{5/7}(1)\)}
                       & \multicolumn{1}{c}{\(\Lambda_{5/7}(10)\)} \\
    \hline
    \textbf{Rotated}   & DL  &4&0.007831~~& 0.007800~~& 6.433~&5.873~ & 0.006782~ & 0.006308~~ &1.894~~ & 1.630~\\
    \textbf{Rotated}   & CAT &5& 0.006551& 0.006494&3.116 & 3.418& 0.005914   &  0.005714 &1.590& 1.488 \\
    \textbf{Rotated}   & GT  &5& 0.006537&0.006411 &3.442  & 3.503 & 0.005878 &  0.005472 & 1.609& 1.411 \\
    \hline
    \textbf{Unrotated} & DL  &4& 0.007648& 0.007676&6.632 &6.225 & 0.007681& 0.007317 &2.957&  2.412 \\
    \textbf{Unrotated} & CAT &5& 0.006396&  0.006276&5.111& 4.559& 0.006321& 0.006026 &2.027 & 1.859 \\
    \textbf{Unrotated} & GT  &5& 0.006649& 0.006491&6.778 & 5.895& 0.006145&0.005727 & 1.946&  1.649 \\
    \hline
  \end{tabular}
  \caption{Combined threshold estimates $p_{\rm th}(\gamma)$ and
    Lambda factors $\Lambda(\gamma)$ for different surface-code
    configurations at boundary noise ratios \(\gamma=1\) and
    \(\gamma=10\). ``Parallel'' and ``across'' refers to orientation of logical
    observables with respect to the interface. $\Lambda_{5/7}$ are
    evaluated at $p_{\rm phys}= 3.45 \times 10^{-3}$.}
  \label{tab:combined_results}
\end{table*}

It is easily seen from Tab.~\ref{tab:combined_results} that both the
threshold error rates and the $\Lambda$-values are significantly
higher when the observable is measured parallel to (rather than
across) the interface.  This is not surprising, since these
observables detect decoding errors corresponding to conjugate logical
operators.  That is, an observable \emph{across} the interface shows
logical errors (domain walls) \emph{along} the boundary.  Likelihood
of such an error is higher even for $\gamma=1$ because of the chosen
geometry; it is further increased for $\gamma>1$ because of higher
fault probability near the boundary.  A quantum code must protect the
encoded data from any errors. Thus, in the subsequent discussion we
focus on the parameters with the observables oriented across the
interface.

It is also seen from Table~\ref{tab:combined_results} that unrotated
codes consistently attain higher thresholds, in agreement with much
higher entropy of minimal-weight error chains in this configuration,
and aligning with well-established findings in surface-code
literature.  This advantage does come with increased qubit overhead,
since unrotated patches require more data qubits to reach a given code
distance.

Unlike for a DL interface, use of CAT and GT gadgets forces an
increase of the bulk measurement circuit depth by 25\%, and one would
expect a comparable decrease of the (pseudo)thresholds with respect to
DL interface.  In reality, at $\gamma=10$, the decrease is about
$13\%$ for rotated surface codes, and close to $18\%$ for regular.  We
believe this to be an entropic effect.  Namely, with a rotated surface
code, multiple errors on the boundary force minimal-weight error
chains to stay close to the interface, thus reducing the entropy and
compensating, to a degree, the threshold reduction.  In contrast,
minimal-weight error chains are linear for regular surface codes, and
their entropy is not expected to change as much.

\textbf{In conclusion}, we performed a careful comparison of three
boundary-connection strategies---DL interface, CAT gadget, and GT
gadget---both for rotated and unrotated surface codes under 
circuit-level depolarizing noise. Our simulations demonstrate
that all three methods maintain fault tolerance even with a
significantly elevated noise at the interface.

DL requires the least resources, adding no ancillary qubits and no
circuit depth beyond standard surface-code measurement cycle.
Respectively, for both rotated and unrotated surface codes, DL
exhibits smallest logical error rates and highest thresholds, which is
not surprising given the absence of additional idling noise.  CAT and
GT gadgets, in contrast, require additional ancillary qubits and an
increase of circuit depth by one CNOT gate.  Nevertheless, both
methods achieve robust error suppression, with the thresholds down
from DL interface values by less that 20\%.  While the thresholds and
the $\Lambda$-factors are consistently smallest for rotated surface
codes, the relative decrease of thresholds and $\Lambda(10)$ from
those for DL interface are both under 10\%.

These results emphasize that boundary gadgets, verified here in
circuit-level quantum memory simulations, can also serve as a basis
for more dynamic approaches, such as lattice surgery in modular
quantum architectures.  Designers must weigh qubit availability,
specialized hardware constraints, gate availability, and error-model
considerations when selecting an interface strategy.  Techniques like
hardware-tailored scheduling, biased-noise optimization, or partial
measurement schemes may further enhance error suppression without
incurring excessive ancillary costs.  Overall, our study confirms that
thoughtfully engineered boundary connections can preserve fault
tolerance in modular surface-code architectures despite considerably
higher interface noise.

{\bf Acknowledgments:} This work was supported in part by the NSF
Division of Physics via the grant 2112848.

\bibliography{lpp,qc_all,more_qc,references}

\onecolumngrid
\appendix


\section*{Appendix}

\begin{figure}[h]
	\centering
	\includegraphics[width=0.75\columnwidth]{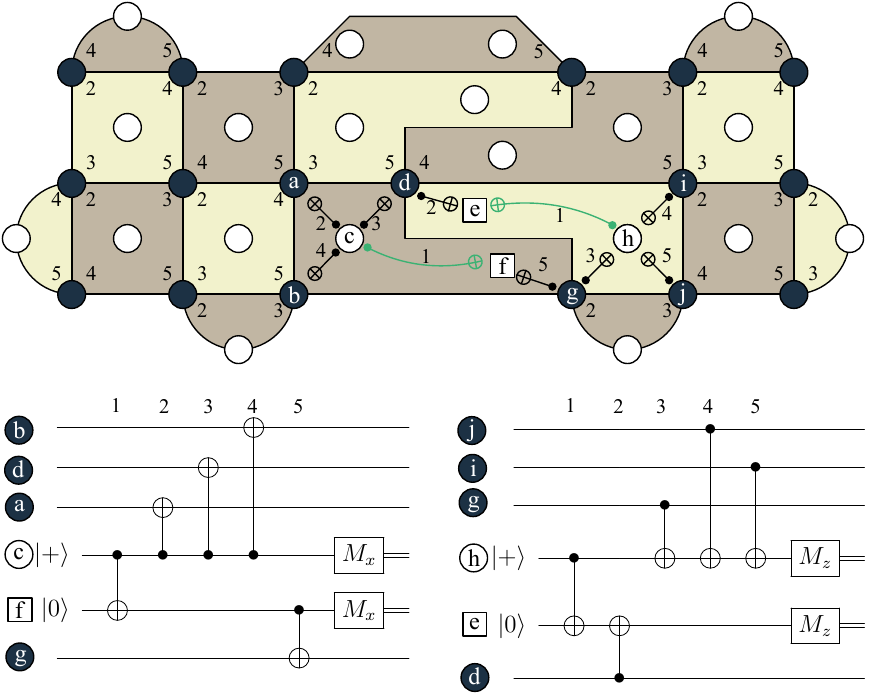}
	\caption{Layout with schedule and circuit for two patches of
          rotated surface code d=3 connecting using CAT. Data qubits
          are dark circles, while ancillas are white circles, the
          squares denotes the added ancillas for CAT. Green CNOT gates
          have elevated noise.}
	\label{Full_Rotated}
\end{figure}

\begin{figure}[h]
	\centering
	\includegraphics[width=0.75\columnwidth]{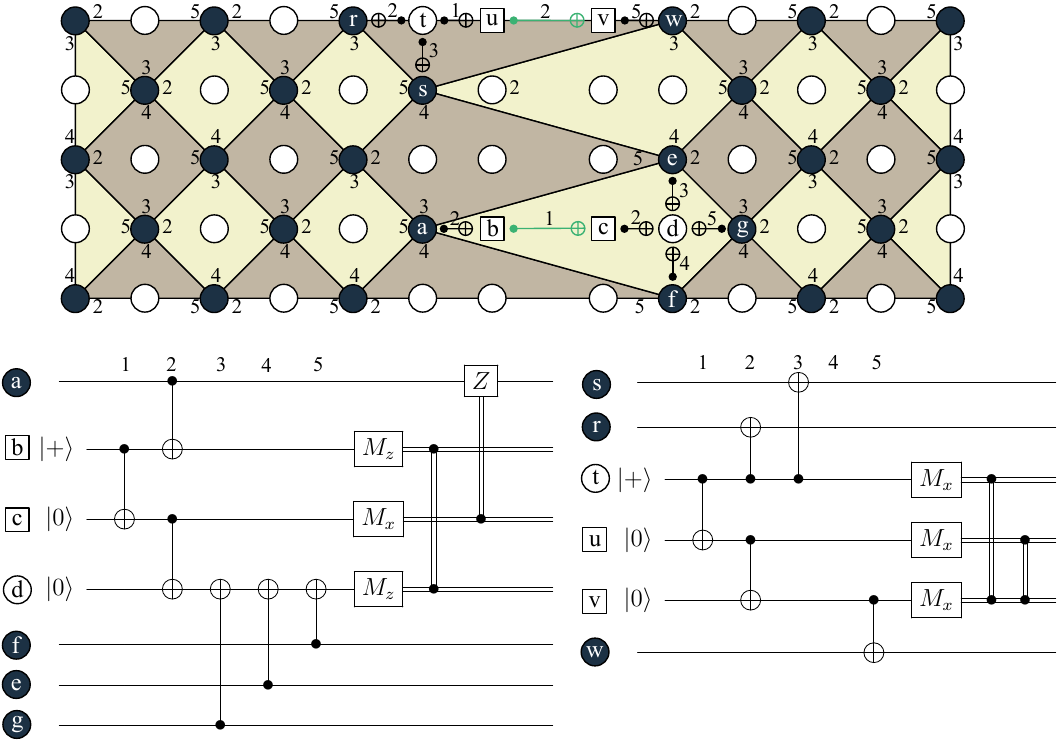}
	\caption{Layout with schedule and circuit for two patches of
          unrotated surface code d=3 connected using GT. Data qubits
          are dark circles, while ancillas are white circles, the
          squares denotes the added ancillas for GT.Green CNOT gates
          have elevated noise.}
	\label{Full_unRotated}
\end{figure}

\begin{figure}[htbp]
	\centering
	\textbf{Rotated Surface Code}\\[0.5em]
	\begin{tabular}{c|ccc}
		& \emph{CAT} & \emph{Gate Teleportation (GT)} & \emph{Direct Link} \\
		\hline \\[-0.8em]
		\rotatebox{90}{\makebox[0.2\textwidth][c]{\emph{Across Observables}}} &
		\includegraphics[width=0.28\linewidth]{Rotated_CAT_Horizontal_gamma10_gamma1} &
		\includegraphics[width=0.28\linewidth]{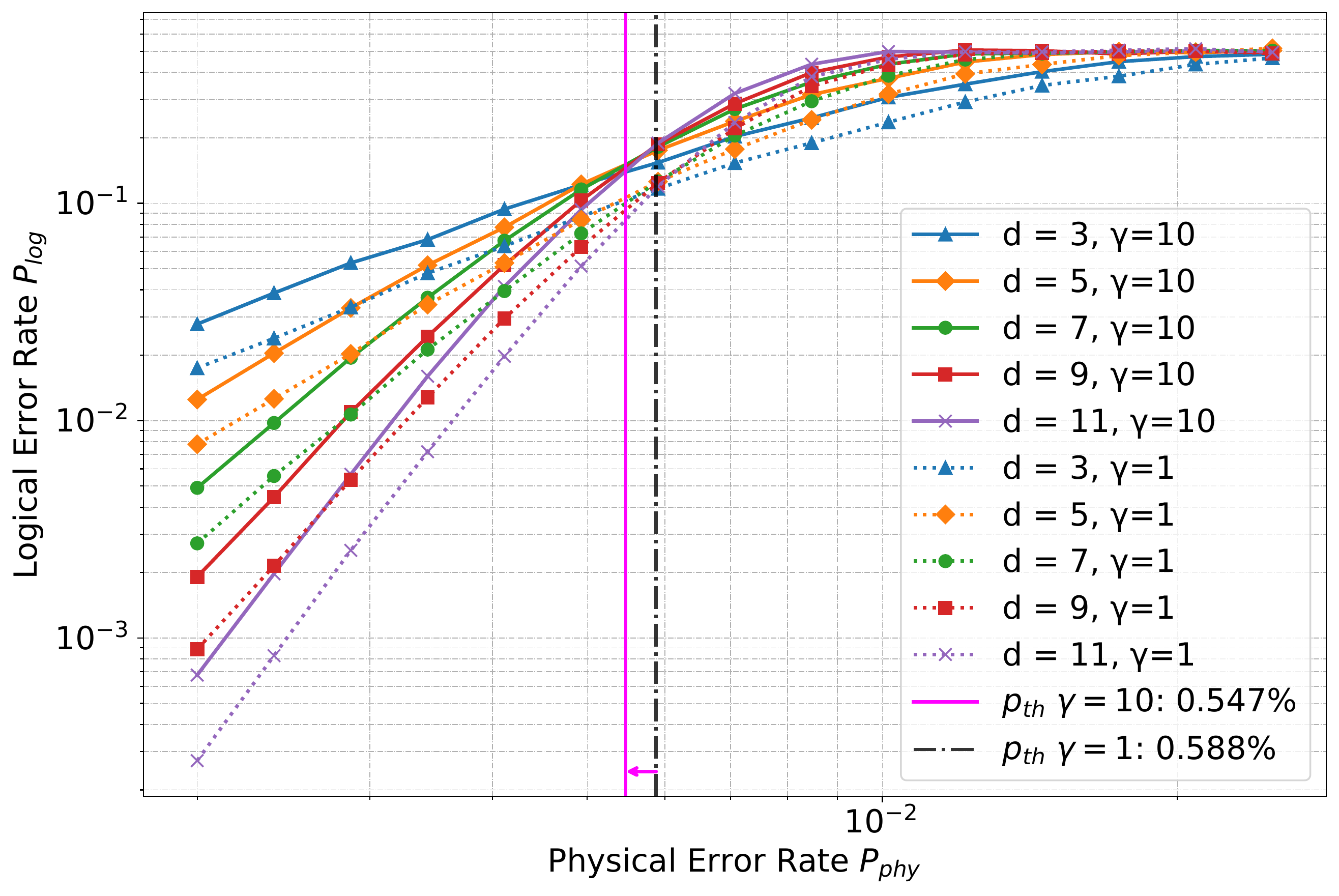} &
		\includegraphics[width=0.28\linewidth]{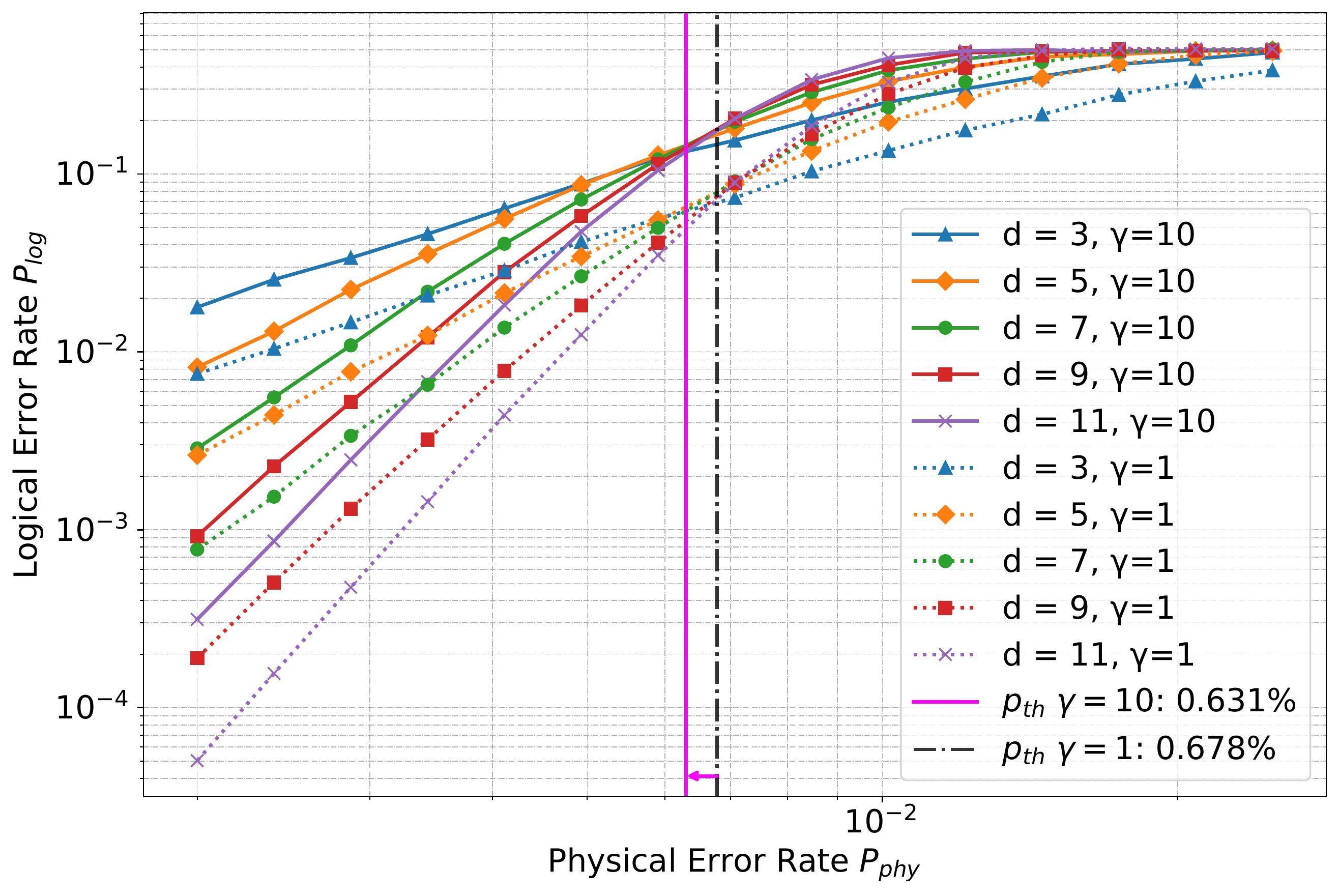} \\
		\rotatebox{90}{\makebox[0.2\textwidth][c]{\emph{Parallel Observables}}} &
		\includegraphics[width=0.28\linewidth]{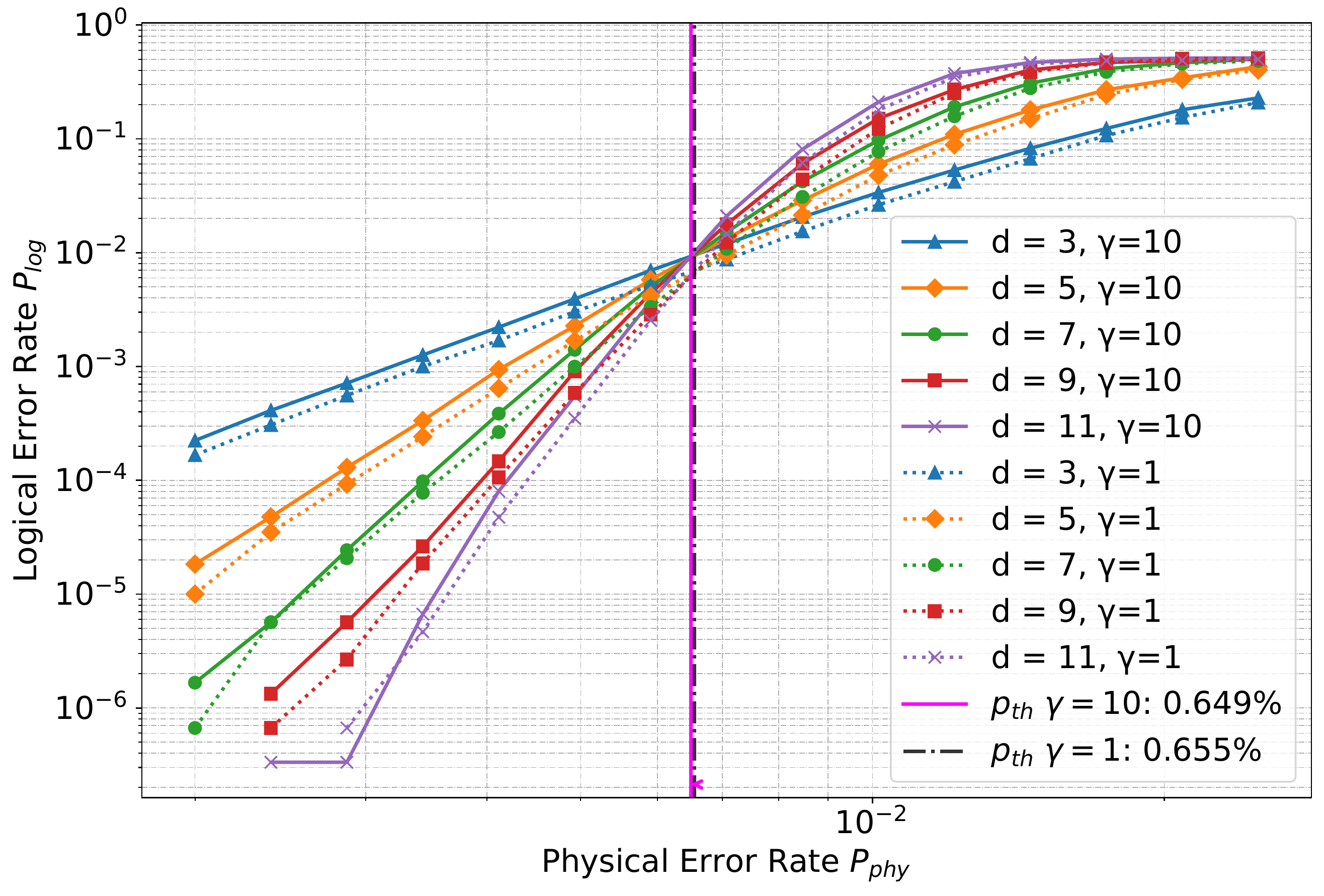} &
		\includegraphics[width=0.28\linewidth]{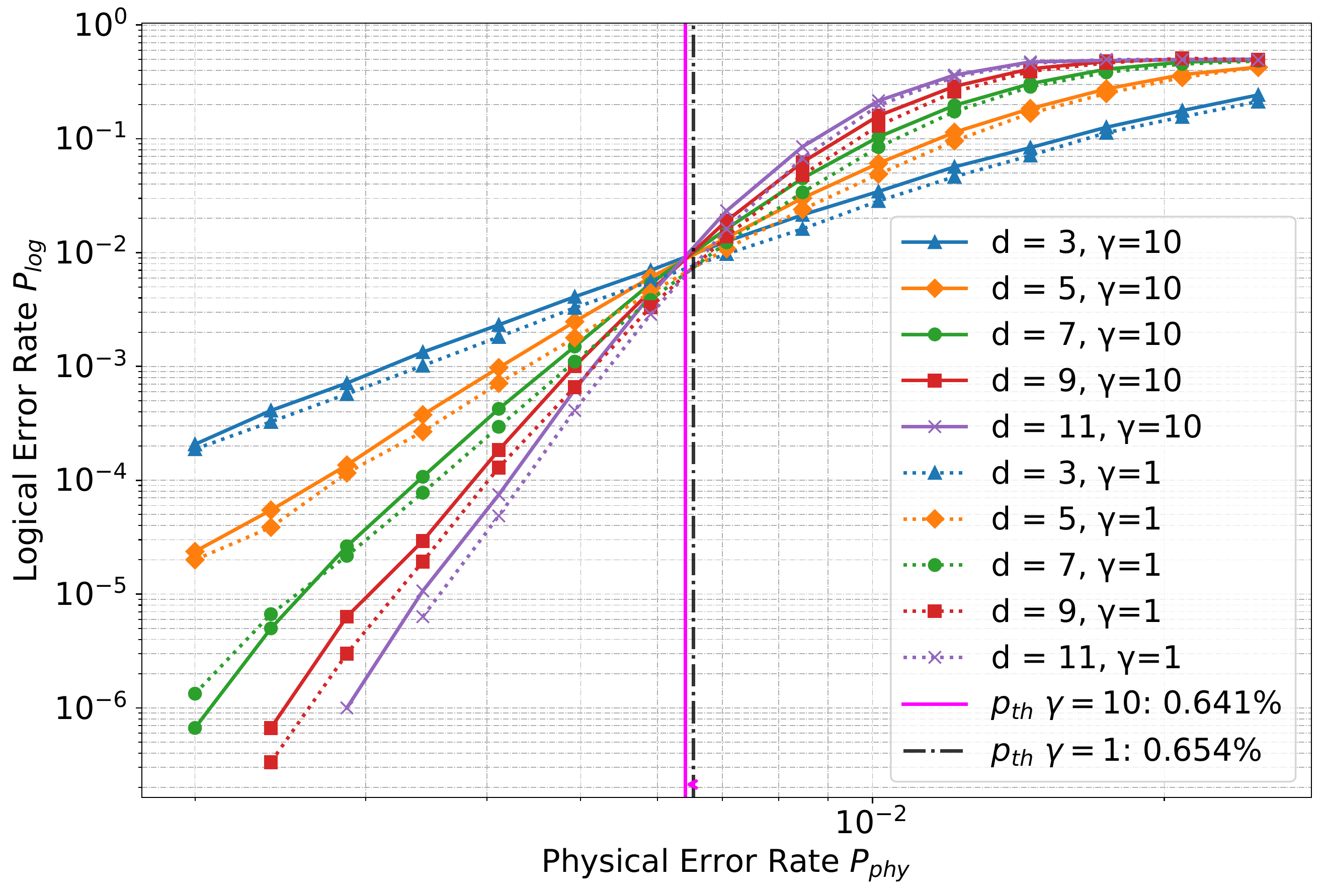} &
		\includegraphics[width=0.28\linewidth]{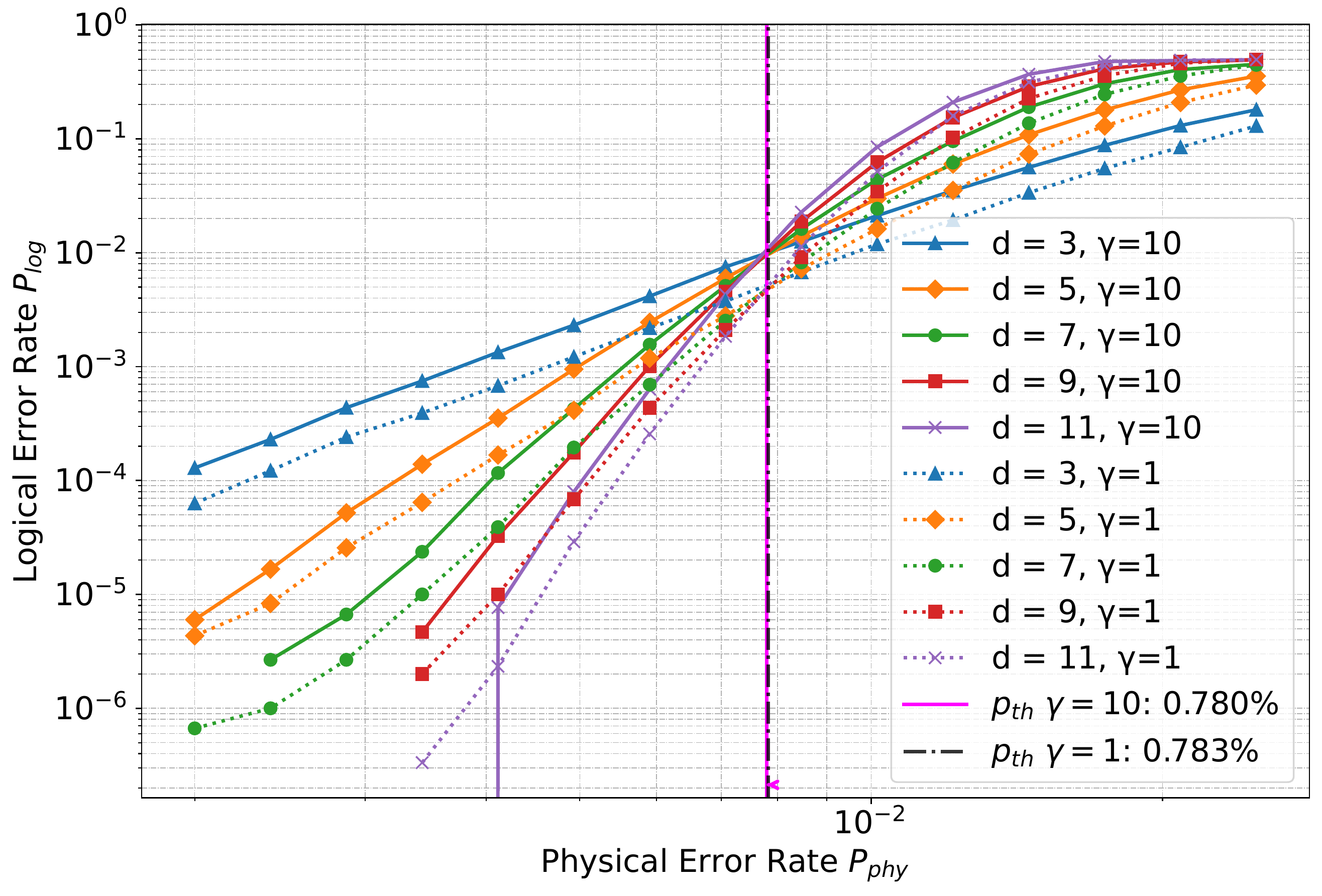} \\
	\end{tabular}

\vspace{1em}

	\textbf{Unrotated Surface Code}\\[0.5em]
	\begin{tabular}{c|ccc}
		& \emph{CAT} & \emph{Gate Teleportation (GT)} & \emph{Direct Link} \\
		\hline \\[-0.8em]
		\rotatebox{90}{\makebox[0.2\textwidth][c]{\emph{Across Observables}}} &
		\includegraphics[width=0.28\linewidth]{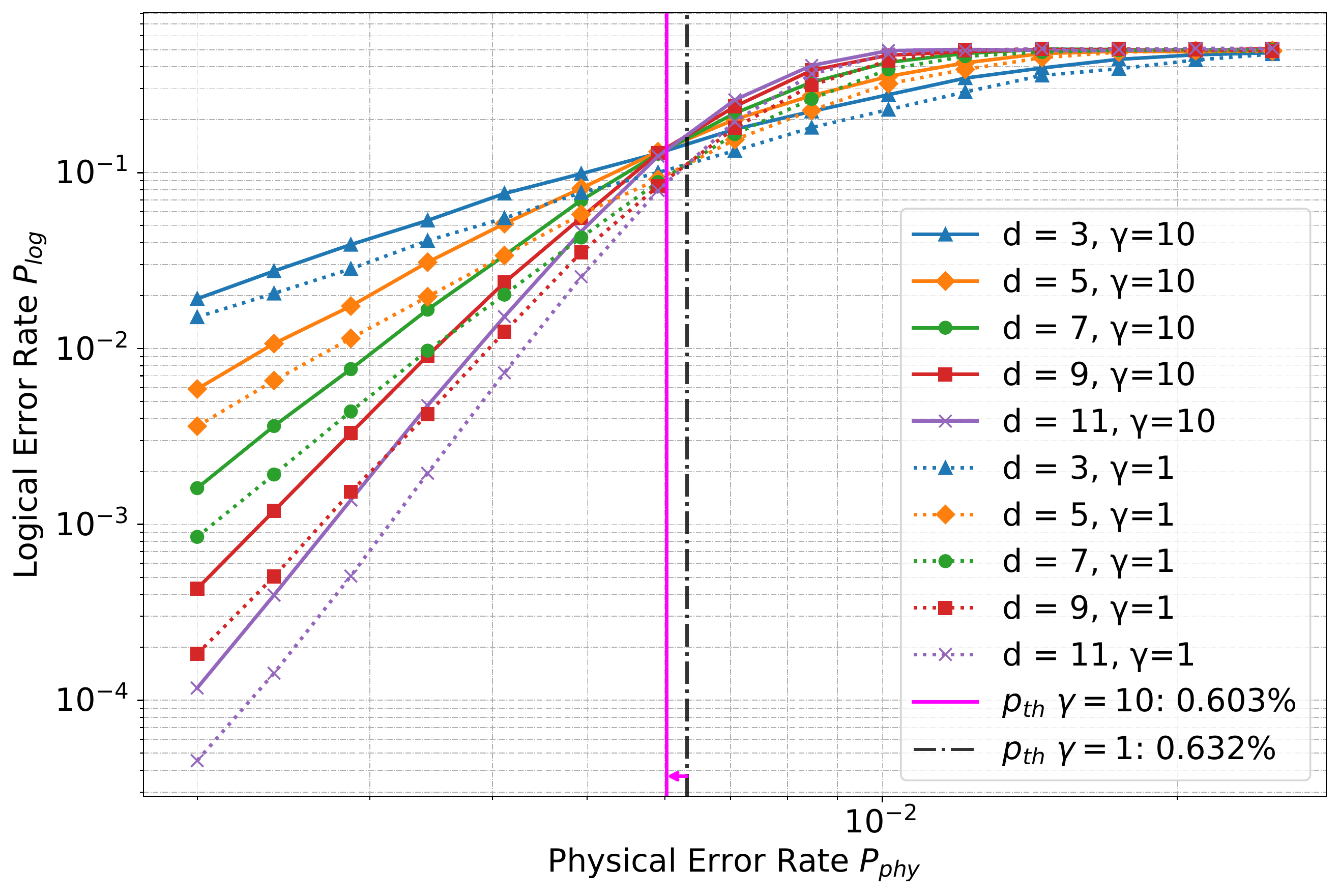} &
		\includegraphics[width=0.28\linewidth]{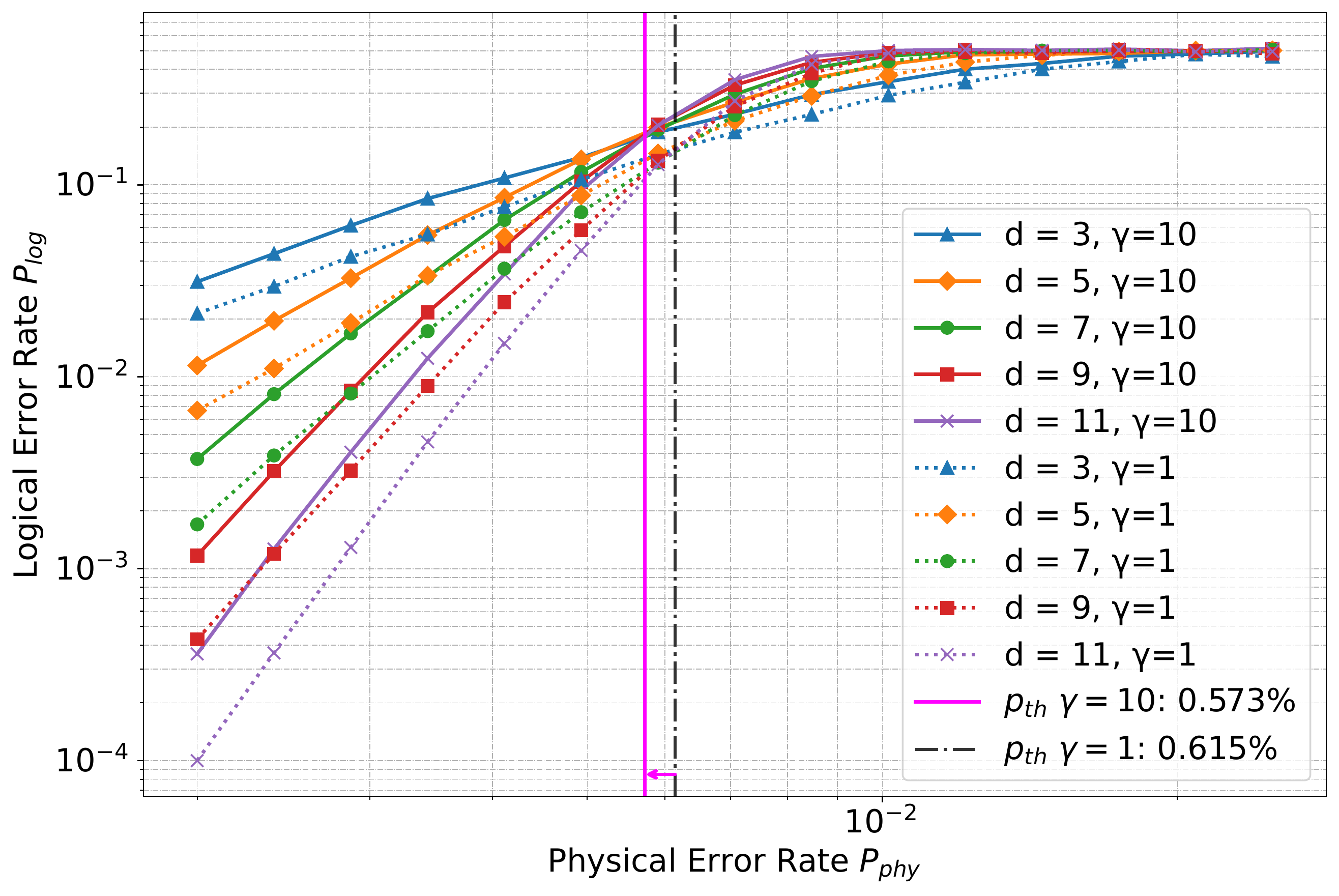} &
		\includegraphics[width=0.28\linewidth]{Unrotated_Direct_Horizontal_gamma10_gamma1} \\
		\rotatebox{90}{\makebox[0.2\textwidth][c]{\emph{Parallel Observables}}} &
		\includegraphics[width=0.28\linewidth]{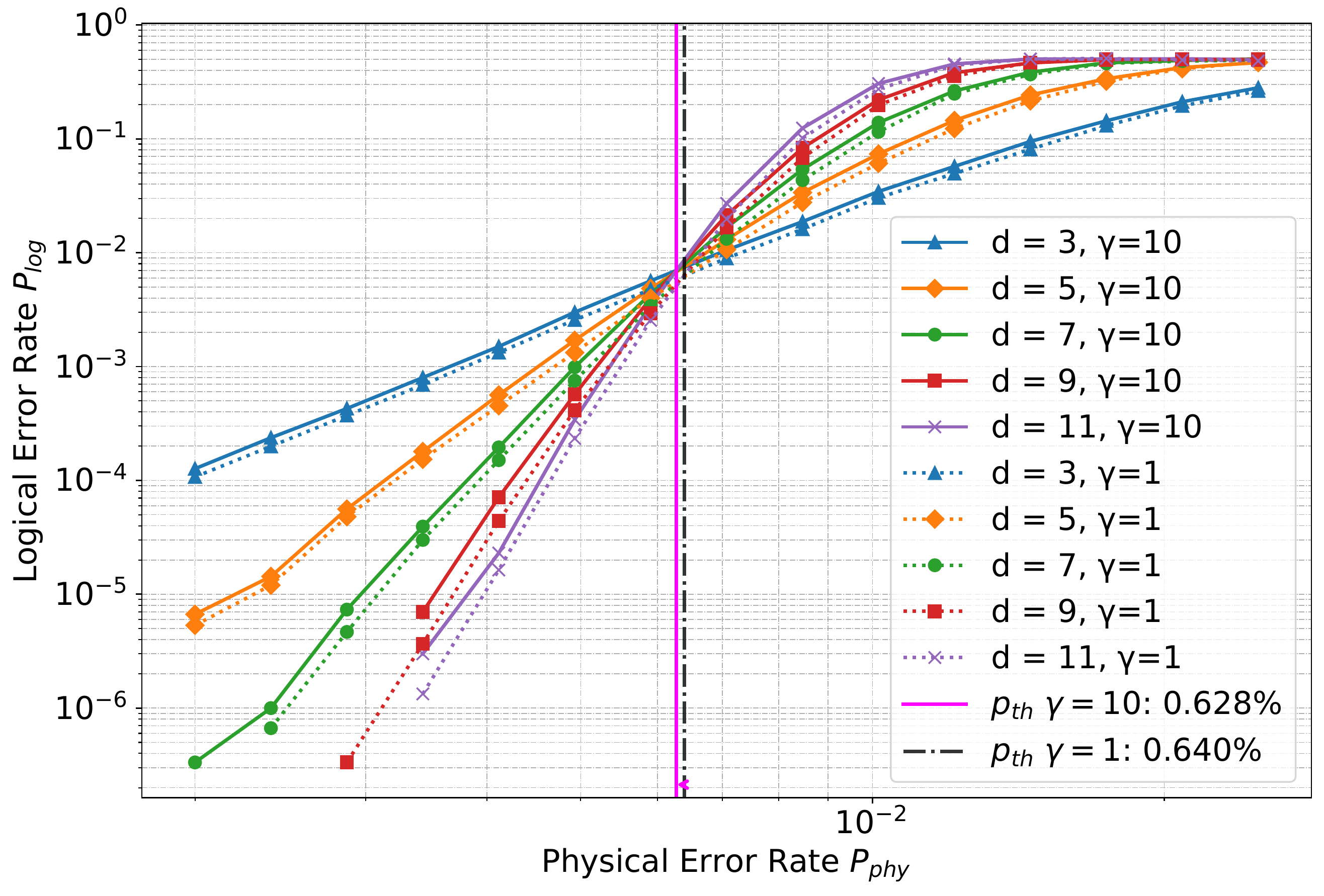} &
		\includegraphics[width=0.28\linewidth]{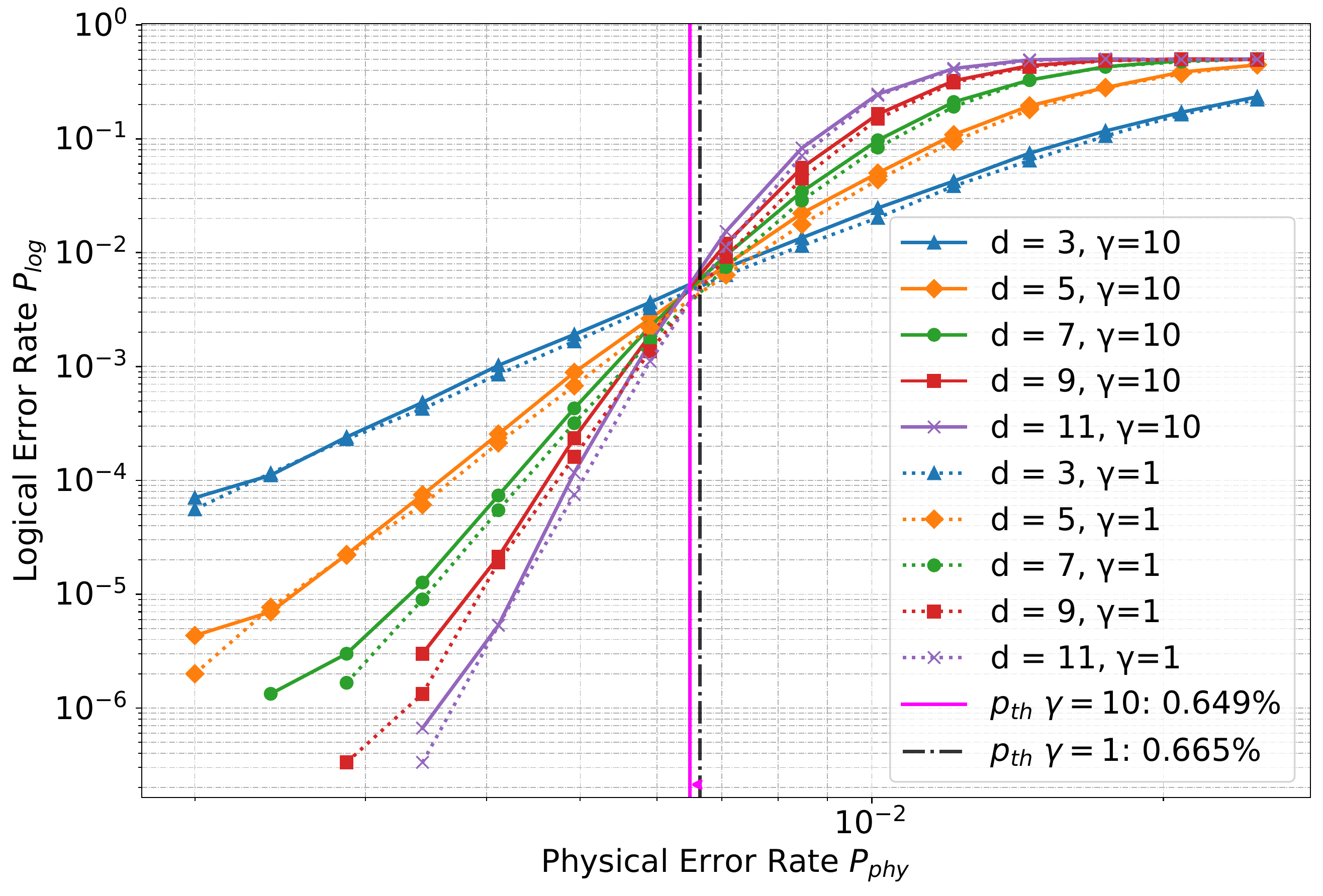} &
		\includegraphics[width=0.28\linewidth]{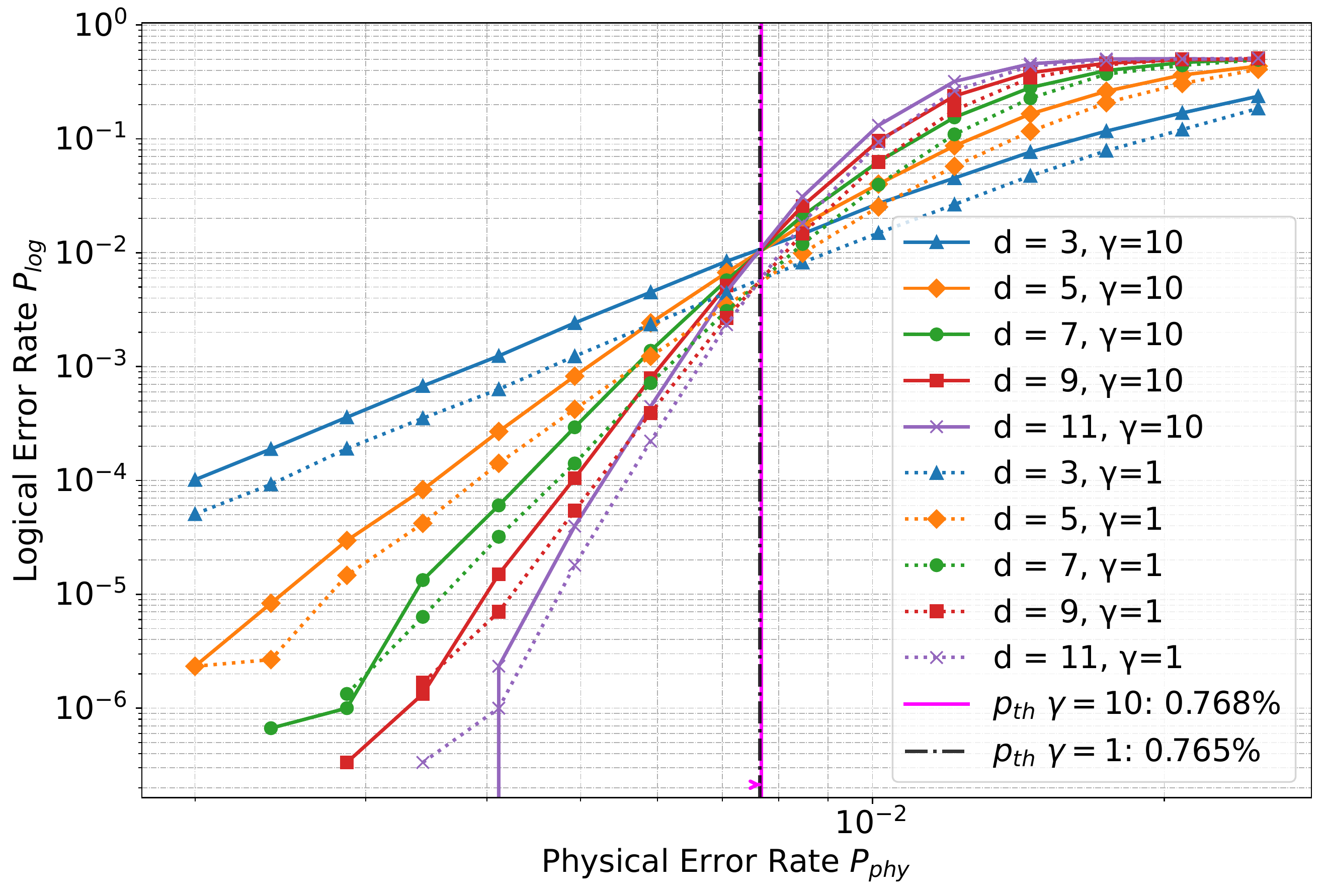} \\
	\end{tabular}
	
	\caption{Comparison of rotated (top) and unrotated (bottom) surface-code configurations at 
		\(\gamma=1\) and \(\gamma=10\). Each row corresponds to logical observables 
		\emph{across} or \emph{parallel} to the interface, while each column shows results for
		a different boundary gadget (CAT, GT, or Direct Link).}
	\label{fig:combined_surface_code_plots}
\end{figure}

Here we give additional details about the constructed circuits and the
threshold-style plots for all interface geometries.  Namely,
Figs.~\ref{Full_Rotated} and \ref{Full_unRotated}, give layouts,
addressing schedules, and full boundary-gadget circuits, respectively,
for rotated surface code patches with CAT gadgets, and unrotated
surface code patches with GT gadgets.
Fig.~\ref{fig:combined_surface_code_plots} shows the complete set of
threshold-style plots grouped by surface code type (rotated or
unrotated), logical observable direction (\emph{across} or
\emph{parallel}), and gadget type [CAT, GT, and DL].  Each plot shows
curves for odd distances in the range $3\le d\le 11$ and for boundary
noise factors $\gamma\in\{1,10\}$.


\end{document}